\documentclass[a4paper, 
12pt, 
oneside]{article} 
\usepackage[english]{babel} 
\RequirePackage[l2tabu,orthodox]{nag} 
\usepackage[utf8]{inputenc} 
\usepackage[T1]{fontenc} 
\usepackage{lmodern} 
\usepackage{upgreek} 
\RequirePackage[babel]{microtype} 
\usepackage{amsmath,amssymb,amsfonts,amsthm,mathtools,wasysym} 
\usepackage{graphicx} 
\usepackage[svgnames]{xcolor} 
\usepackage{marginnote} 
\usepackage{tabularx} 
\usepackage{csquotes} 

\usepackage[margin=1in]{geometry}

\usepackage[backend=bibtex, 
block=none, 
maxbibnames=4, 
backrefstyle=three, 
arxiv=abs, 
sorting=none, 
firstinits=true, 
style=numeric-comp, 
doi=false, 
url=false, 
backref, 
hyperref, 
isbn=false, 
defernumbers=true]{biblatex} 

\usepackage[bookmarksopen, 
bookmarksdepth=3]{hyperref} 
\hypersetup{bookmarksnumbered, 
linktocpage, 
colorlinks, 
debug, 
citecolor=[rgb]{0,.7,.1}, 
urlcolor=[rgb]{.3,.7,0}
} 

 % ... 

\renewbibmacro{in:}{}
\AtBeginBibliography{%

}
\AtEveryBibitem{%
      \clearfield{eprintclass}%
}

\DeclareFieldFormat{doilink}{%
\iffieldundef{doi}{#1}%
{\href{http://dx.doi.org/\thefield{doi}}{\color{blue}#1}}} 
\DeclareBibliographyDriver{article}{%
  \usebibmacro{bibindex}%
  \usebibmacro{begentry}%
  \usebibmacro{author/translator+others}%
  \setunit{\labelnamepunct}\newblock
  \usebibmacro{title}%
  \newunit
  \printlist{language}%
  \newunit\newblock
  \usebibmacro{byauthor}%
  \newunit\newblock
  \usebibmacro{bytranslator+others}%
  \newunit\newblock
  \printfield{version}%
  \newunit\newblock
  \usebibmacro{in:}%
  \printtext[doilink]{%
  \usebibmacro{journal+issuetitle}%
  \newunit
  \usebibmacro{byeditor+others}%
  \newunit
  \usebibmacro{note+pages}%
  }%
  \newunit\newblock
  \iftoggle{bbx:isbn}
    {\printfield{issn}}
    {}%
  \newunit\newblock
  \usebibmacro{doi+eprint+url}%
  \newunit\newblock
  \usebibmacro{addendum+pubstate}%
  \setunit{\bibpagerefpunct}\newblock
  \usebibmacro{pageref}%
  \usebibmacro{finentry}}

\DeclareSymbolFont{extraup}{U}{zavm}{m}{n}
\DeclareMathSymbol{\varheart}{\mathalpha}{extraup}{86}
\DeclareMathSymbol{\vardiamond}{\mathalpha}{extraup}{87}

\newcommand{\bea}{\begin{eqnarray}}
\newcommand{\eea}[1]{\label{#1}\end{eqnarray}}
\newcommand{\be}{\begin{equation}}
\newcommand{\ee}{\end{equation}}
\newcommand{\bitm}{\begin{itemize}}
\newcommand{\eitm}{\end{itemize}}
\newcommand{\bmat}{\begin{pmatrix}}
\newcommand{\emat}{\end{pmatrix}}
\newcommand{\bal}{\begin{aligned}}
\newcommand{\eal}{\end{aligned}}
\newcommand{\ba}{\begin{align}}
\newcommand{\ea}{\end{align}}
\newcommand{\bat}[1]{\begin{alignat}{#1}}
\newcommand{\eat}{\end{alignat}}
\newcommand{\bse}{\begin{subequations}}
\newcommand{\ese}{\end{subequations}}
\newcommand{\beq}{\begin{equation}}
\newcommand{\eeq}[1]{\label{#1}\end{equation}}

\newcommand{\rf}[1]{(\ref{#1})}

\newcommand{\cev}[1]{\reflectbox{\ensuremath{\vec{\reflectbox{\ensuremath{#1}}}}}}

\newcommand{\remark}[1]{\vspace{5pt}\noindent\textbf{Remark} : #1\vspace{5pt}} 
\renewcommand{\footnote}[1]{\textcolor{red}{\footnotemark}\footnotetext{\,#1}} 

\renewcommand{\d}{\text{d}}

\renewcommand{\H}{\textrm{H}}

\newcommand{\tspin}{\tfrac{1}{2}}

\newcommand{\pgh}{\textrm{pgh}\hspace*{1pt}}
\newcommand{\agh}{\textrm{agh}\hspace*{1pt}}
\newcommand{\gh}{\textrm{gh}\hspace*{1pt}}

\newcommand{\aghn}{\textrm{agh}\hspace*{1pt}\#}
\newcommand{\ghn}{\textrm{gh}\hspace*{1pt}\#}

\newcommand{\E}{\textrm{E}}

\newcommand{\vertex}[2]{#1\hspace*{-2pt}-\hspace*{-2pt}#2\hspace*{-2pt}-\hspace*{-2pt}#2}

\def\Ds{\displaystyle{\not{\!D\!\,}}}
\def\As{\displaystyle{\not{\!\!A\!\,}}}

\def\ds{\displaystyle{\not{\!\partial\!\,}}}

\newcommand{\tdim}{\textit{\tiny D}}

\newcommand{\tintspace}{\hspace*{-1pt}}

\def\a{\alpha}

\def\G{\Gamma}
\def\D{\Delta}

\def\ve{\varepsilon}

\newcommand{\Slash}[1]{{\ooalign{\hfil#1\hfil\crcr\raise.167ex\hbox{/}}}}

\addto\captionsenglish{
  \renewcommand{\contentsname}%
    {}%
}

\begin{document}

\begin{center}
~\\~\\~\\
\noindent
{\fontsize{22pt}{10pt} \textbf{The Elegance of Cohomological Methods}} \\
\vspace*{5pt}
{~} \\
\vspace*{1pt}
\noindent
{\fontsize{15pt}{10pt} \textbf{A Physicist's Approach to the BRST-Antifield Formalism}}\,\footnote{To be submitted to \emph{Reviews of Modern Physics}.}
~\\~\\~\\~\\~\\

{\Large Gustavo \textsc{Lucena Gómez}}

~\\~\\~\\

{\em Albert Einstein Institute, Am M\"{u}hlenberg 1, Golm, D-14476 Germany}\\
\vspace*{5pt}
{\em Institute of Physics AS CR, Na Slovance 2, Prague 8, Czech Republic}
\end{center}
~\\

\begin{center}
\textbf{Abstract}
\end{center}

\noindent
The BRST-Antifield reformulation of the deformation problem is reviewed in a self-contained and heuristic way. The focus is on finding all consistent interaction terms for fields propagating on a Minkowskian $D$-dimensional spacetime. Particular emphasis is put on the physical interpretation for the mathematical objects of the formalism. \\

\vspace*{10pt}
\begin{center}
------------------
\end{center}
\vspace*{-30pt}
\tableofcontents

\newpage 

\begin{center}
\textbf{Disclaimer}
\end{center}
\noindent
This review article contains nothing original besides, perhaps, its presentation. It is short and offers a physicist's perspective, and leaves out many demonstrations and historical considerations. However the constructive approach that is followed hopefully renders the material somewhat accessible. Noticeably, the list of references is far from being exhaustive. Any type of remark regarding the present notes would be much appreciated. 

\section*{Introduction}\addcontentsline{toc}{section}{Introduction}

A typical problem in Theoretical Physics is the following: given a free field theory for some field content, find out which interaction terms are allowed by consistency requirements such as gauge invariance. We refer the unfamiliar reader to Section~\ref{context} where we briefly recall the interpretation of gauge theories and the usual way they are dealt with. 

Given an interaction term it is in principle straightforward to check whether it is gauge invariant or not. However, from a constructive viewpoint one would wish to start from some field content or, equivalently, from some free action principle, and proceed to an exhaustive classification of all possible interaction terms. Such a task is far from being straightforward, mainly for the following two reasons. First, the freedom of field redefinitions together with that of integration by parts implies that the same interaction term can be cast into seemingly different forms. Second, looking for interaction terms which are gauge invariant under the \emph{free} gauge transformation rules is not the most generic thing to do. The \emph{free} gauge transformations are those that leave the free theory invariant, but an interacting theory which is not invariant under the free gauge transformations may nevertheless be invariant under some \emph{deformed} version of those transformations. In brief the problem of adding interaction terms to a free Lagrangian is difficult because one asks two questions at the same time and seeks a common answer, that is, one looks for vertices that deform the Lagrangian and possibly the gauge transformations at the same time in a consistent manner, and moreover field redefinitions must be taken into account.  

The traditional way of dealing with this deformation problem is via the Noether procedure: suppose one finds a tentative vertex, to be added to some free theory. If the said vertex is non-Abelian, then it deforms the original, Abelian gauge transformations. One needs to verify whether the deformed gauge transformations form some deformed gauge algebra or not. If they do, then one needs to determine the algebraic structure thereof, and the procedure needs to be repeated for every putative vertex. The BRST-Antifield reformulation\footnote{Also called BRST-BV. We use all such terminologies interchangeably.} of the deformation problem allows for proceeding \emph{backwards}: the formalism makes it very natural to start with the classification of the possible deformations of the gauge algebra. The advantage is that the possible deformations of the gauge algebra are much constrained, and in the BRST-BV language those constraints translate to precise cohomological statements, so that one is assured to be exhaustive. One then follows a systematic procedure called the \emph{consistency cascade}, which consists in solving cohomology equations in order to determine whether the tentative algebra deformation is allowed or not. The obstructions to consistency are again related to precise cohomology classes, and if the algebra deformation is consistent the procedure also yields the corresponding gauge-symmetry and Lagrangian deformations.

The BRST-BV reformulation cleverly deals with the two difficulties outlined above: field redefinitions and deformation of the gauge transformations. Instead of using the standard (free) action a (free) \emph{master action} is constructed which on top of the standard action contains terms with explicit information about the gauge transformations. The deformation problem then translates to that of deforming the master action, thus naturally dealing at the same time with the problem of deforming the Lagrangian as well as the gauge transformations in a consistent way. Also, a BRST operator is constructed which implements at the same time the gauge transformations and the field redefinitions, and relates them to precise cohomology classes. It is thus ensured that the freedom granted by field redefinitions is fully used when classifying different gauge-invariant vertices. 

All these advantages come at a price, namely that of having to introduce auxiliary fields: \emph{antifields} and \emph{ghosts} (and \emph{antighosts}), which enlarge the original phase space (also referred to as the configuration space). The philosophy is that in this enlarged phase space, to be defined below, it is possible to reformulate statements and properties such as gauge invariance and on-shell triviality into precise cohomological equations.\footnote{Recall that a quantity is said to be on-shell zero, or on-shell trivial, if it can be made so by making use of the equations of motion.} In particular, the usual properties of consistency and non-triviality will be related to the two familiar aspects of cohomological calculus: computing the kernel of a nilpotent operator as well as the trivial part therein. In a first approach, the formalism presented here may therefore seem excessive, but we nevertheless think it not only extremely useful but also quite natural. Furthermore the BRST-BV reformulation of the deformation problem is intrinsically off-shell in spirit, that is, no gauge-fixing is required nor implied. 

We owe the original BRST--BV formalism to Becchi, Rouet, Stora, Tyutin as well as to Batalin and Vilkovisky \cite{Becchi:1974md,Becchi:1975nq,Tyutin:1975qk,Batalin:1981jr}, which they originally developed in order to address the quantization of gauge theories. Later on it was realized that one could use this language also at the classical (i.e. non-quantum) level to consistently search for deformations of given gauge theories \cite{Barnich:1993vg,Barnich:1994db,Barnich:1994mt,Barnich:1994cq}. This is the application we present, in the simplified case where one deforms a free theory. The literature on the BRST-Antifield reformulation of the deformation problem includes the very good review \cite{Henneaux:1997bm} and we also point out the algebraic and geometrically oriented lectures \cite{Henneaux:1989jq}, the report \cite{Barnich:2000zw} as well as the comprehensive book \cite{henneauxbook}, all of which go beyond the scope of the present review. 

\section{Contextualization of the Problem for Non-Experts}
\label{context}

In broad strokes, the evolution in time (dynamics) of particles of some type is described by means of a field theory for a corresponding set of fields. Mathematically speaking the fields are sections of some fiber bundle over a base manifold which represents the spacetime the particles propagate on. In this review we consider particles propagating on a Minkowski spacetime of dimension $D\geq 4$, so that the base manifold is $\mathbb{R}^D$ equipped with the Minkowski metric $\eta_{\mu\nu}$ ($\mu,\nu = 0,\dots, D-1$).\footnote{Our universe seems to be approximately Minkowskian and to have four (large) spacetime dimensions. Note that mathematically any pseudo-Riemannian manifold can be taken as the base manifold.} The isometry group of this manifold is the $D$-dimensional Poincaré group, containing the Lorentz group as a subgroup, and the fields can be considered concretely as tensors for the latter, that is, they transform under irreducible representations of the Lorentz group. These fields depend on spacetime coordinates which parametrize the base manifold, one of which is the time, the others being called space coordinates. 

Constructing a field theory for such fields means constructing an action $S$, which is a functional of our spacetime-dependent tensor fields, and which contains all the information about the theory in a way that we will detail soon. The action, or \emph{action principle}, is the integral of a density over the base manifold. The density is called the \emph{Lagrangian}, and many questions in Field Theory can be reformulated in terms of it. As all the rest of the present work is about deforming Field Theory Lagrangians in some set-up by means of the BRST-Antifield techniques, let us explain succinctly the way they are understood and dealt with in high-energy physics. Note that what follows is a short and sketchy contextualization summarizing standard Field Theory textbook material, and for a more mathematically oriented presentation of Field Theory we refer to \cite{Deligne:1999qp}. 

More precisely, the aforementioned fields correspond to the type of particles one wishes to consider. The prescription of (classical) Field Theory is that one chooses a vacuum value for these fields and then considers perturbations (equivalently, excitations) of the fields around this vacuum. These excitations are really what the particles are associated with, and they are tensors for the whole Poincaré group of isometries of spacetime. The time evolution of these particle fields is determined by solving the Euler--Lagrange equations, which are obtained by equating to zero the functional variation of the action with respect to the field excitations. These \emph{equations of motion}, as they are called in physics, are partial differential equations which are solved by specifying a number of independent functions of the space coordinates at some given time, namely a certain number of functions on the Cauchy surface corresponding to that given time.\footnote{We are eluding a number of subtleties related to solving equations of motion.} In physics speak it is declared that the corresponding particle contains or propagates that number of degrees of freedom.\footnote{Truly what determines the number of degrees of freedom is the number of independent (quantum) numbers one needs to specify in order to determine the (quantum) state of the particle completely. This does not always match the number of independent solutions to the free equations of motion, for reasons outside the scope of this introduction.} 

In some descriptions the number of components of a field excitation does not match the particle's number of degrees of freedom. Such redundant formulations are called \emph{gauge theories}, and the fields they describe transform under \emph{gauge symmetries} accounting for the mismatch. For example, the gauge invariant description of the photon is by means of an action depending on the Maxwell one-form field $A = A_\mu(x)\d x^\mu$, transforming in the fundamental (vectorial) representation of the Lorentz symmetry group. On a Minkowski spacetime of dimension $4$ the photon propagates $2$ degrees of freedom, whereas the vector field $A_\mu(x)$ evidently has $4$ components. The mismatch of two is corrected by gauge invariance: it is postulated that the relevant object is not the field $A_\mu(x)$ but, rather, the equivalence class of such fields, where in this case the equivalence relation reads $A \to A + \d\lambda(x)$. It is beyond of the scope of this introduction to work out the details of this example, but it is a standard exercise in Field Theory for which we refer e.g. to \cite{henneauxbook}. 

Such gauge-invariant formulations might be perceived as being overly complicated. Why not deal with some set of components corresponding to the number of degrees of freedom carried by the particle of interest\,? Such a formulation is called a (completely) \emph{gauged-fixed} one, or a formulation without gauge invariance. However, for many purposes having a gauge-invariant formulation of the theory is more practical. One can always gauge-fix a given gauge theory, thereby obtaining its gauge-fixed version, and it is straightforward to do so. The converse, however, is not easy in general, and gauge-invariant formulations are usually thought of as being harder to obtain in general. 

Dealing with fields transforming under some gauge symmetry\footnote{We shall stick to the standard language lore according to which gauge transformations are called gauge symmetries. It should be clear, however, that those are not true symmetries of our theory\;---\;no physical statement can be drawn from the knowledge that our theory is gauge invariant\;---\;, unlike e.g. the symmetries leaving the relevant spacetime manifold invariant, which do have physical consequences.} comes at a price: the action functional needs to be gauge invariant. If it were not, it would mean that our action, the object containing all the physical information about our theory, depends on the field representative chosen within its gauge equivalence class. Differently put, our action would depend on the gauge choice, i.e. it would be unphysical. All physical quantities need be gauge invariant. When trying to construct an action for a given set of fields, the requirement of gauge invariance can be a hard one to meet. On a more philosophical level, let us note that this is not necessarily a 'price' that we pay. Indeed, physicists often look for guiding principles, constraining rules that may restrict the number of theories one may a priori think of. A modern view is that gauge invariance, although it is \emph{not} a physical thing but a mere redundance in the way we formulate our theory, is nevertheless a guide of great value for building physical theories. \\

An action can be either \emph{free} or \emph{interacting}. What this means at the conceptual level is that a free action is one which describes the simplified scenario where particles freely propagate in some spacetime, without interacting with other particles or with themselves. They do not decay, collide, or exchange energy in any way\;---\;they just propagate. An interacting action is one which describes the general but more intricate situation in which we consider interactions among the different particles. How is the difference understood at the mathematical level\,? The action is the integral of some density over the spacetime manifold, and the density is a polynomial functional in the fields and their spacetime derivatives. The density, also called the Lagrangian starts at degree $2$ in the fields and generically contains higher-order pieces. The degree-$2$ part is the free part of the action, also called the kinematical part, whereas the higher-degree terms are referred to as the interaction \emph{vertices}, or interaction terms. As the various equations of motion are obtained by equating to zero the functional variation of the action with respect to the various fields, free equations of motion are linear in the fields. 

Given a set of fields it is straightforward to construct the free action describing their propagation over Minkowski spacetime. For example, in the case of a scalar field $\phi(x)$ (transforming according to the trivial representation of the Poincaré group), we immediately see that the only possible kinetic, free density is
\begin{equation}
\partial_\mu \phi(x) \partial^\mu \phi(x)\,.
\end{equation}
The Einstein convention is used according to which repeated indices are contracted with the Minkowski $D$-dimensional metric $\eta_{\mu\nu}$ and its inverse $\eta^{\mu\nu}$ (conventions are given in the main part of the text). A scalar field does not transform under any gauge symmetry, and as we said it does not transform under the Lorentz group either because it belongs to the trivial representation thereof. As a kinetic term needs at least one spacetime derivative, the above is the simplest term we can think of, since it has to be Lorentz invariant (which is why the two derivative indices need to be contracted). One could argue that both the derivatives could act on the same field $\phi$ instead, but recalling that the ultimate object is the spacetime integral of the above we see that, neglecting boundary terms,\footnote{Boundary terms is a very important aspect of dealing with actions. However, their treatment is far beyond the scope of this work and we will limit ourselves to pointing out that, locally in spacetime, the dynamics may be described without taking them into account provided that appropriate boundary conditions are chosen for the fields.} such combinations are equivalent thanks to the freedom of performing integrations by parts. We will not consider kinetic terms with more than two spacetime derivatives, and the above expression is unique up to those.  

The upshot is that free theories are easy and may be taken as the starting point of a program aiming at constructing an interacting action principle for some set of fields. Why is it not equally easy to add to free Lagrangians interaction terms of higher degree in the fields\,? The answer has much to do with gauge invariance. Lorentz invariance also constrains the possible terms one can think of, but that requirement is straightforwardly dealt with: if all spacetime indices are contracted Lorentz invariance holds. Gauge invariance is more difficult to check, and even more so to enforce a priori. The above example of the scalar field is blind to the problem because the field does not transform under a gauge symmetry (the gauge equivalence class is trivial). Thus any functional of the scalar field and its derivatives is gauge invariant and constructing interaction vertices is easy: $\phi\phi\phi$, $\phi\phi\phi\phi$ or other such combinations are all gauge invariant interaction terms. Let us then go back to the example of the vector field $A_\mu(x)$, whose kinetic term reads
\begin{equation}
\d A \wedge \star\,\d A \propto (\partial_\mu A_\nu \partial^\mu A^\nu - \partial_\nu A_\mu \partial^\mu A^\nu)\,\d^\textsc{d}x\,,
\end{equation}
where $\star$ denotes the Hodge dual of the curvature $\d A$. One can think of the following interaction vertex: $V = A_\nu A^\nu A_\mu A^\mu $. However, it is easy to check that such is not an invariant combination under $A \to A + \d \lambda(x) = A + \delta A$. That is, one computes the gauge variation $\delta V$ of $V$, using the chain rule, and observes that it is nonzero for a generic $\lambda(x)$, even up to total derivatives. 

\section{BRST and Antifields: Getting Started}
\label{sec:brst}

We restrict ourselves to the case where the theory one starts with is free, but in principle the same framework can be used to address the problem of deforming a theory which is not free~---~see e.g. \cite{Henneaux:1989jq}. Another assumption we make is that the free gauge theory is \emph{irreducible}.\footnote{Irreducibility of the theory means that the gauge transformations are mutually independent \cite{henneauxbook}.} This means that we start from a free, irreducible gauge theory of a collection of fields $\{\phi^i\}$, with $m$ Abelian gauge invariances
\be 
\label{gaugetransfrecall}
\delta_{\ve}\phi^i \equiv R^i_\a\ve^\a, \quad \a=1,2,\dots,m ,
\ee
which leave the free action $S^{(0)}[\phi^i]$ invariant. The $R^i_\a$ usually are differential operators. We present the formalism in generic dimension~$D$. We do so in the following seven steps:
\begin{description}
\item[Step 1:] Ghosts for Gauge Parameters,\vspace*{-5pt}
\item[Step 2:] Antifields and Gauge Variations, \vspace*{-5pt}
\item[Step 3:] Longitudinal Differential Along Gauge Orbits, \vspace*{-5pt}
\item[Step 4:] Koszul--Tate Differential, \vspace*{-5pt}
\item[Step 5:] BRST Operator, \vspace*{-5pt}
\item[Step 6:] Consistency Cascade,\vspace*{-5pt}
\item[Step 7:] Second-Order Consistency and Antibracket.
\end{description}

In Step~$1$ and $2$ we enlarge the phase space of our original fields $\phi^i$. Step~$1$ replaces gauge parameters by ghosts, which are added to the configuration space, whereas Step~$2$ introduces \emph{antifields}, which source gauge transformations in some generalized action called the \emph{master action}. Then, in Step~$3$ and $4$ we reformulate two important concepts in cohomological terms. The first one, dealt with in Step~$3$, is gauge invariance, and it will connect with Step~$1$. The second, addressed in Step~$4$, is field redefinitions, and it will relate to Step~$2$. Step~$5$ then combines both reformulations of these concepts into a single, unified operator: the BRST differential, conveniently implementing both the equations of motion and the gauge symmetries. Once the correct operator has been identified, in Step $6$ we explain how to search for consistent interactions in this formalism, that is going through the \emph{consistency cascade}, which allows one to start from potential deformations of the gauge algebra. Finally, Step~$7$ is concerned with second-order consistency and quartic vertices. In the latter we discuss the \emph{antibracket}; a symplectic structure on our enlarged phase space which not only allows for an easier analysis of second-order consistency but also for a `geometrical' reformulation of the deformation problem in general, which we touch upon briefly. 

\subsection*{Step 1: Gauge Parameters for Ghosts}\addcontentsline{toc}{subsection}{Step 1: Gauge Parameters for Ghosts}

The first step is as follows: each gauge parameter $\ve^\a$ is replaced by a corresponding ghost field $\mathcal C^\a$, so that the gauge transformations \rf{gaugetransfrecall} now read
\be 
\label{gaugetransfrecallghosts}
\delta\phi^i \equiv R^i_\a \mathcal{C}^\a, \quad \a=1,2,\dots,m\, .
\ee
The ghost $\mathcal C^\a$ is declared to have the same algebraic symmetries but
opposite Grassmann parity as $\ve^\a$. This means that, e.g. if some gauge parameter is bosonic (as for example that of a spin-$1$ gauge field), the corresponding ghost is Grassmann odd, and vice versa. 

\remark{we do not spell out spacetime indices, and the index $\alpha$ in $\mathcal C^\a$ is accounting for the various ghosts replacing the various gauge parameters. As no assumption is made on the spin of the fields and their associated gauge parameters, it should be kept in mind that each of the latter can have spacetime indices, but we treat them all generically. When we say that the ghosts have the same algebraic symmetries as the original fields they correspond to, we are referring to those spacetime indices, and mean that both transform in the same representation of the Lorentz group.}

The ghosts are now included in the phase space, thus enlarging that of the original fields, and all of them are sometimes collectively also called fields, which we denote by $\{\Phi^A\} \equiv \{\phi^i, \mathcal{C}^\a \}$. In order to keep track of the nature of each of the fields we further introduce a grading, called the \emph{pure ghost number}, defined to be $0$ for the original fields and $1$ for the ghosts: 
\begin{subequations}
\begin{align}
\pgh (\phi^A) &\equiv 0 \,,\\
\pgh (\mathcal{C}^\a) &\equiv 1 \,.
\end{align}
\end{subequations}

\subsection*{Step 2: Antifields and Gauge Variations}\addcontentsline{toc}{subsection}{Step 2: Antifields and Gauge Variations}

Let us define the free \emph{master action}, corresponding to the original, free action $S^{(0)}[\phi^i]$. As we anticipated in the beginning of the present section, the point is to build a generalized action which, in addition to containing information about the original Lagrangian will also contain explicit information about the gauge transformations.\footnote{We use the word \emph{explicit} for the following reason: given some standard (free) action one can always work out the corresponding gauge symmetries, so that this information is already contained in the action functional, although in an implicit manner.} It is such a free master action, denoted $S_0$ (note the subtle change in notations), which is deformed later on. The idea is the following: in addressing the deformation problem for non-Abelian vertices, one is not only looking for deformations of the Lagrangian invariant under the original gauge transformations. Rather, one needs to allow for the gauge transformations to get deformed too. One virtue of the master action is that it contains explicit information about both these aspects, and the deformation problem, when formulated in terms of it, will automatically take into account both these features in a way which ensures consistency and exhaustivity. The free master action is defined as follows:
\be 
S_0 \equiv S^{(0)}[\phi^i] + \int\d^\tdim x \,\phi^*_iR^i_\a\mathcal C^\a \,,
\ee
where the $\phi^*_i$ are the \emph{antifields}, which play the role of sources (i.e. Lagrange multipliers) for the gauge variations in the master action. 

The configuration space is even further enlarged by also introducing antifields corresponding to the ghosts, named antighosts. Collectively we thus have antifields $\Phi^*_A$, which are again defined to have the same algebraic properties as the corresponding $\Phi^A$ but opposite Grassmann parity (which correctly makes the above master action Grassmann even). Our phase space is now given by $\{\Phi^A, \Phi^*_A\}$, where $\{\Phi^*_A\} \equiv \{\phi^*_i, \mathcal{C}^*_\a \}$. Note that in Step $1$, the gauge parameters are \emph{replaced} by ghosts and those are then added to the phase space. Here, rather, we \emph{supplement} the phase space with antifields corresponding to the fields. \\

Note that the antighosts do not enter the above master action, and at this stage one can think of them as being added also for the sake of democracy~---~their role will be clarified in the sequel. As for the antifields, besides sourcing the gauge variations in the master action, in Step $4$ we shall see that they have another role to play. However for the moment let us be content with the explicit presence of information about the gauge symmetries in the master action. The rule will be that whatever multiplies the antifields in the master action is the gauge transformation of the corresponding field. This will be of much use when deforming the free theory. 

\remark{a word of caution about the interpretation of the above action should be added. For the unfamiliar reader, it might be tempting to consider the antifields and the ghosts as auxiliary fields in the usual sense of Field Theory. The corresponding paradigm is that the equations of motion which follow from variating the action with respect to those auxiliary fields allow one to solve for them, hence integrating them out when plugging their algebraic expressions (in terms of the dynamical fields) back into the action. However, such is not the way in which one should understand the master action. Rather, the latter is really a tool, allowing to keep gauge invariance and field redefinitions under control, and should not be thought of as a standard action.}

Finally, as we have introduced new (anti-)fields we need a new grading to keep track of who is who in the phase space of all the fields and antifields. We thus define the \emph{antighost number} as
\begin{subequations}
\begin{align}
\agh (\Phi^A) &\equiv 0\,, \\
\agh (\Phi^*_A) &\equiv \pgh(\Phi^A) + 1\, .
\end{align}
\end{subequations}
Also, we need to extend the definition of the pure ghost number to the antifields, and the correct definition is $\pgh(\Phi^*_A) \equiv 0$. 

\subsection*{Step 3: Longitudinal Differential Along Gauge Orbits}\addcontentsline{toc}{subsection}{Step 3: Longitudinal Differential Along Gauge Orbits}

As explained at the beginning of the present section, this third step has to do with Step $1$. What is done is to define an operator $\Gamma$ implementing the (free) gauge variations on our enlarged phase space. The definition is the following:
\begin{align} 
\label{gammaghosts}
&\Gamma \phi^i \equiv R^i_\a \mathcal{C}^\a \,, & \Gamma\mathcal{C}^\a \equiv 0\,,
\end{align}
that is, $\Gamma$ is the `longitudinal derivative along the gauge orbits' \cite{henneauxbook}. From the above definition one sees that $\Gamma$ is Grassmann odd and that $\Gamma^2 = 0$. Note that the latter property, or equivalently $\Gamma\mathcal{C}^\a \equiv 0$, reflect the fact that the gauge algebra of the free theory is Abelian \cite{henneauxbook}. To finish implementing the gauge variations one needs to further define the action of $\Gamma$ on the antifields and the antighosts, and the correct definition is $\Gamma \Phi^*_A \equiv 0$.  

As for every nilpotent operator (of degree two), it is natural to consider the \emph{cohomology} of $\Gamma$ $\H (\Gamma)$, 
\be 
\H (\Gamma) \equiv \{X \in \textrm{phase space} \;|\; \Gamma X = 0, \, X \sim X + \Gamma Y\}\, ,
\ee
where $X$ and $Y$ are local expressions built out of the fields and antifields, and `$\sim$' is the equivalence relation. This equivalence relation means that we consider $X$ and $X + \Gamma Y$ as the same object. Put in mathematical terms, an element in $\H(\Gamma)$ is not a local expression involving fields and antifields but, rather, a \emph{class} of such expressions, where two combinations $X$ and $X'$ belong to the same class if and only if there exists some $Y$ such that $X = X' + \Gamma Y$, i.e. if and only if they are equivalent to one another by virtue of the equivalence relation defined hereabove. Another way of putting it is that the cohomology is the set of $\Gamma$-closed quantities \emph{up to} $\Gamma$-exact quantities, or \emph{modulo} such exact quantities. The various terminologies will be used interchangeably. Note that we will sometimes abuse the correct terminology of classes, and simply speak about a particular object belonging to the cohomology or not, but one should always remember that two objects differing by exact terms are only `counted once' when considering the cohomology. 

The physical interpretation is as follows: the cohomology of $\Gamma$ is the set of gauge-invariant combinations, up to gauge variations of something else (up to pure gauge objects, one might say). With this definition we are really beginning our formalization of the properties that are crucial for us. Indeed, all the above definition does is formalize the definition of what `being an observable' means~---~to be refined in Step $5$. We cannot possibly stress enough that this approach is at the core of the present reformulation, and in fact it is precisely the idea of associating physical quantities with cohomological classes of nilpotent operators which the BRST-BV approach put forward \cite{Becchi:1974md,Becchi:1975nq,Tyutin:1975qk}. Note that $\Gamma$ has pure ghost number equal to $1$ but leaves the antighost number unchanged. \\

Before moving on to Step $4$ we should add a few words about terminology and notation. In the language of cohomology, a combination which is annihilated by some operator is said to be \emph{closed}, and one that can be expressed as the application of the operator to some other quantity is said to be \emph{exact}. An equivalent way of phrasing things is to declare any closed object a \emph{cocycle} and any exact one a \emph{coboundary}. We shall switch back and forth between both terminologies, although we prefer the former. The cohomology of $\Gamma$, for example, will be said to be the space of all $\Gamma$-closed combinations which are not $\Gamma$-exact. Also, a $\Gamma$-exact element shall sometimes be said to be \emph{trivial} in the cohomology, and therefore we shall sometimes drift towards the standard abuse of terminology according to which `being in the cohomology' is understood as being $\Gamma$-closed and $\Gamma$-exactness is expressed as being trivial in $\H (\Gamma)$. 

When we shall solve for the first-order deformation of the free master action by requiring it to be gauge invariant (up to field redefinitions), those terms proportional to the antifields will represent the deformation of the gauge symmetries, whereas the quantities containing the original fields only are deformations of the original, free Lagrangian. As for terms containing the antighosts, much like the terms multiplying antifields in the master action are the gauge symmetries, the terms multiplying the antighosts shall be seen to correspond to deformations of the gauge algebra. This fits quite nicely with the fact that the free master action $S_0$ contains no such terms, for the algebra of the free theory is Abelian. 

\subsection*{Step 4: Koszul--Tate Differential}\addcontentsline{toc}{subsection}{Step 4: Koszul--Tate Differential}

Step $3$ had to do with Step $1$, and the present step has to do with Step~$2$, that is with the antifields. Indeed, so far we have dealt with gauge invariance, and we further need to address the freedom granted by field redefinitions. Again, this will be done by the introduction of an odd operator, this time named $\Delta$. The possibility of field-redefining our deformations may sound like one which can be handled easily even in the standard approach, but it is not, and although one may like to think of $\Gamma$ as being the main ingredient of the ultimate BRST differential $s = \Gamma + \Delta$ (see Step $5$), the inclusion of $\Delta$ is in fact crucial. It will ensure that field redefinitions are not left unconsidered when passing to the cohomology of $s$, describing our gauge invariant observables. \\

In order to properly deal with fields redefinitions we again follow the fruitful paradigm according to which `physical' combinations should be associated with elements in the cohomology of our nilpotent operator $\Delta$. More precisely, this means that the quantities we wish to consider as redundant (defining some equivalence class) should be associated with trivial elements of the cohomology. The analogy with the previous step, concerned with gauge invariance, is thus clear: in the same way as $\Gamma$-exact objects are gauge variations of something else, which are the redundancies corresponding to the possibility of performing gauge transformations (see Section~\ref{context}), $\Delta$-exact objects will be associated with combinations which are on-shell zero, that is, with field redefinitions.\footnote{Note that $\Delta$-exact terms will describe all types of field redefinitions, including expressions stemming from redefinitions of the gauge parameters.} We thus would like to define the action of $\Delta$ on our phase space in a way such that the equations of motion for our original fields are equal to $\Delta$-variations of something else. The correct definition is
\be 
\label{eomsdelta}
\Delta \phi^*_i \equiv \E_i (\phi^j) \,,
\ee
where $\E_i$ is defined to be the equation of motion for $\phi^i$, that is, $\delta S^{(0)} \equiv \E_i  \delta \phi^i$ (where $\delta$ here denotes a functional variation with respect to the fields). Moreover, starting from the above definition, the Bianchi identities (zeroth-order Noether identities) corresponding to the equations of motion are easily seen to enforce the following relations:
\be 
\Delta (R^i_\a \phi^*_i ) = 0\,. 
\ee
It can then be shown that having objects such as $R^i_\a \phi^*_i$ in the cohomology of $\Delta$ leads to inconsistencies in the formalism \cite{Henneaux:1989jq,henneauxbook}. However the cure to this problem is obvious: if an object is $\Delta$-closed and we wish to exclude it from the cohomology, the way out is to make it $\Delta$-exact. Accordingly, we define
\be 
\Delta C^*_\a \equiv R^i_\a \phi^*_i  \,.
\ee
Our antighosts finally play a role\,! Note that, from the relation \rf{eomsdelta} one concludes that $\Delta$ is also Grassmann odd, just like $\Gamma$, and its nilpotency is again guaranteed (and can be checked explicitly on every field). One may think of the above relation in the following way: we are just adding the antighosts to the game in order to express $R^i_\a \phi^*_i$ as a $\Delta$-exact quantity, much in the same way as we simply added the antifields because we needed an object that could source the gauge variations in the master action (see Step $2$).\\

The action of the $\Delta$ operator on the rest of the original fields is derived by acting with it on \rf{eomsdelta}, which yields $\Delta \phi^i = 0$, and the action on the ghosts then follows from applying $\Delta$ to \rf{gammaghosts} and noticing that $\Delta$ anticommutes with $\Gamma$, for they are both Grassmann odd. We are ready to formulate our second cohomology, that of $\Delta$:
\be 
\H (\Delta) = \{X \in \textrm{phase space} \;|\; \Delta X = 0, \, X \sim X + \Delta Y\}\, .
\ee
The physical meaning is, again, quite clear: the cohomology of $\Delta$ is the space of all quantities built out of the $\Phi^A$ which are not themselves $\Delta$-variations of something else, that is quantities which are not proportional to the equations of motion (equivalently, which are not field redefinitions, or on-shell zero). We are making progress towards a complete and refined formalization of what we mean by an observable: a non-trivial, gauge invariant quantity identified with others up to field redefinitions and gauge variations. Evidently, the correct operator which will compute for us the `physical' cohomology will need to combine both $\Delta$ and $\Gamma$. That operator is called the BRST differential, and is formally introduced in Step $5$ here below. 

Regarding the gradings, $\Delta$ has $\agh (\Delta) = -1$ and $\pgh (\Delta) = 0$, as is easily deduced from its action on the various fields and antifields. 

\subsection*{Step 5: BRST Operator}\addcontentsline{toc}{subsection}{Step 5: BRST Operator}

With the above considerations in mind we can finally construct our ultimate nilpotent operator: the BRST differential $s$. The aim is that its cohomology should correctly describe the notion of an observable. Differently put, $\H(s)$ should be associated with the space of inequivalent, gauge-invariant and non-trivial deformations $S_1$ of the free master action $S_0$ to be elucidated below. The definition which computes the correct cohomology is the following:
\be 
s \equiv \Gamma + \Delta \,,
\ee
and one can again check its nilpotency, either by noticing that $\Gamma$ and $\Delta$ anticommute or directly. Its cohomology,
\be 
\H(s) = \{X \in \textrm{phase space} \;|\; s X = 0, \, X \sim X + s Y\} \,,
\ee
is exactly the one computing all the consistent action deformations up to field redefinitions. Indeed, it is the space of combinations which are on-shell gauge invariant but which are not themselves the gauge variation of something else or a field redefinition of something else, as follows from simply investigating the cohomology conditions in light of the decomposition $s = \Gamma + \Delta$. \\ 

Let us now consider a consistent deformation of the free master action into some deformed master action $S$, that is,
\be 
\label{masteractionpert}
S = S_0 + gS_1 + g^2 S_2 + \cdots\, ,
\ee
where the ellipses stand for higher order deformations. In the generic case where the deformation is possibly non-Abelian, the gauge transformations also get deformed in such a perturbative way, and $S_0$ is assumed to be invariant under the zeroth-order gauge transformations. We shall primarily address the problem of first-order deformations. Now, as it is well known and also easy to check, for the first order piece $S_1$ the requirement of perturbative gauge invariance is really that of being invariant under the \emph{free} gauge symmetries~---~this is equally true whether one considers the master action or the original action. Thanks to this fact, that we have implicitly used until now, we can analyze the problem of finding and classifying the consistent deformations by means of the BRST differential $s$, which implements all at once the necessary requirements of (free) gauge invariance and (free) on-shell triviality. In cohomological terms the condition our first-order deformation of the free master action must satisfy reads
\be 
S_1 \in \H (s) \,.
\ee

Finding all the consistent first-order deformations up to equivalence is thus tantamount to computing the cohomology of $s$, which is a well-defined mathematical problem. However, we shall now proceed to introducing one last refinement of the cohomology we want to compute, and that is the one of partial integration. Indeed, we shall always assume that the deformations we seek are local,\footnote{The fact that locality is compatible with the formalism is proved in \cite{Henneaux:1991zz}.} that is, they are spacetime integrals of functionals of our phase space variables (the fields, antifields, \dots) and of derivatives thereof, provided the derivatives appear up to finite order only. Our notation goes
\be 
S_1 \equiv \int \tintspace a \,.
\ee
Therefore our problem can be and will be reformulated in terms of $a$, namely, at the level of the (master) Lagrangian instead. Consequently, provided we are interested in the local dynamics only, to which boundary terms in the action never contribute, we have the freedom of performing integrations by parts. The relevant cohomology is thus not $\H(s)$ but, rather, the cohomology of $s$ \emph{modulo $\d$}, noted $\H(s | \d)$, which is defined as $\H(s)$ but with the extra freedom of performing partial integrations. The ultimate condition that our deformation must satisfy then reads
\be 
a \in \H (s | \d) \,,
\ee
which is both necessary and sufficient. On top of this condition, our deformation $a$ might of course be required `by hand' to preserve certain global symmetries, such as Lorentz invariance or parity. Note that the above condition can also be rewritten as $a \in \H (s + \d)$.\\

Before we can start analyzing the above condition in detail there is one more grading we need to introduce, namely the \emph{total ghost number} (or simply the \emph{ghost number}), equally defined on all fields as the pure ghost number minus the antighost number. That we do so now is not an accident. Indeed, the BRST differential $s$ does not have neither definite pure ghost number nor definite antighost number, as is inferred from the properties of $\Gamma$ and $\Delta$. The correct quantum number which keeps track of the action of $s$ is the (total) ghost number, and in fact we find $\gh (s) = 1$. The ghost number of the various fields and antifields are straightforwardly computed, and also given in Table~$1$ at the end of the section, together with the action of the various operators on the various fields and antifields.  

With our last quantum number, the total ghost number at hand, it is time we mention a condition on the deformation we have been neglecting so far, and it is that of the quantum numbers which it must have. Firstly, let us note a simple yet important fact: just like $s$ the free master action does not have definite pure ghost or antighost number, but it has total ghost number $0$, which the reader shall easily verify. Note that this further indicates that the BRST differential is the right operator to consider when deforming the free master action but, more importantly, it means that $S_1$ and $a$ must have total ghost number $0$. This will much restrict the possible ingredients one may use to build a tentative deformation term. Also, as we shall see the elements of $\H(s|\d)$ with higher ghost number shall also enter the game at some point. The subset of $\H(s|\d)$ having ghost number equal to $k$ is called the cohomology at ghost number $k$ and it is denoted by $\H^k(s|\d)$. The final word about the deformation is thus the following:
\be 
\label{acohom}
\boxed{\vphantom{\tspin}\;a \in \textrm{H}^0(s | \d)\;} \, .
\ee
This condition repackages all the requirements of consistency and non-triviality of the deformation.

\subsection*{Step 6: Consistency Cascade}\addcontentsline{toc}{subsection}{Step 6: Consistency Cascade}

Having formulated in a precise way the cohomology we wish to compute, we now explore how to do so in a clever way. An obvious thing to do is to use the gradings we have introduced to further inspect the problem. It is found that the antighost number is most useful in doing so, and the main reason for that is the following theorem: let
\be 
a = a_0 + a_1 + a_2 + \cdots \,,
\ee
where $\agh (a_i) \equiv i$ (note negative antighost numbers cannot occur). The theorem, proved in \cite{Barnich:1994mt} under very generic assumptions, states that $a_i = 0 \;\forall \, i > 2$. This result is, in general, very strong and as we shall see below it will be crucial in being able to analyze the deformation in a systematic way. Let us point out that for cubic deformations the theorem is trivially proved, in the sense that there is no $\ghn = 0$ combination of three of our fields and antifields of antighost number higher than $2$, as one can directly observe by considering the various quantum numbers we have assigned each of the fields. In the sequel we shall confine ourselves to cubic deformations. 

\remark{as we are addressing first-order deformations anyway, the reader might wonder what it means to further confine ourselves to cubic deformations. Could one think of e.g. quartic first-order deformations\,? Although this situation never arises in physics it is nevertheless a logical possibility. In \cite{Boulanger:2008tg} it has been proved that such deformations never occur up to spin $5$ and argued to be true for all spins. In the present approach we shall not prove the analogous result and simply make the assumption that our first-order deformations are cubic.}

The above result is more useful than it might seem at first glance. First of all, the interpretation of the three pieces appearing in the above decomposition of $a$ is much clear: $a_0$ is the deformation of the Lagrangian, $a_1$ is the deformation of the gauge symmetries and $a_2$ is the deformation of the gauge algebra\,! This can be verified by noticing that $a_0$ contains only the original fields ($\aghn = 0$), $a_1$ contains one original field, one ghost and one antifield ($\aghn = 1$) and $a_2$ contains two ghosts and one antighost ($\aghn = 2$)~---~and once again recalling that $a$ has total ghost number zero. 

Secondly, with the above decomposition in mind the cohomology condition 
\be 
\label{as}
s\hspace*{0.5pt}a + \d (...) = (\Gamma + \Delta)(a_0 + a_1 + a_2) + \d (...) = 0\,,
\ee
when analyzed antighost number by antighost number, gives rise to three independent conditions:
\begin{subequations}
\label{cocycle}
\begin{align}
\G a_2&\doteq 0\,,\label{cocycle1}\\
\D a_2+\G a_1&\doteq 0\,,\label{cocycle2}\\
\D a_1+\G a_0&\doteq 0\,.\label{cocycle3}
\end{align}
\end{subequations}
These conditions form what is known as the \emph{consistency cascade}, and we have used a new notation: `$\doteq$' is understood as the standard equality up to total derivatives. Moreover, another general theorem \cite{Barnich:1994mt} teaches us that one can always assume $\Gamma a_2 = 0$, which is stronger than the same condition up to total derivatives. 

The above consistency cascade will be our main tool in finding out consistent deformations, and it is worth commenting on. The equation \rf{cocycle3} for $a_0$ is familiar: it expresses the fact that the deformation of the Lagrangian, $a_0$, is invariant up to field redefinitions $\Delta a_1$ and total derivatives. The two remaining equations are less easily interpreted, but their role is to ensure first-order consistency of the deformation of the Lagrangian. Intuitively, the situation is clear: \rf{cocycle2} involves $a_1$ and $a_2$, and is thus ensuring that the deformation of the gauge symmetries induced by $a_0$ closes to a gauge-algebra deformation $a_2$. Then, Equation \rf{cocycle1} ensures consistency of the gauge-algebra deformation $a_2$. In fact, one can check that \rf{cocycle2} is the first-order projection of the condition that the gauge symmetries close to some algebra and \rf{cocycle1} is a consistency condition for the gauge-algebra deformation, again projected to first order in the deformation. \\

Having established the above consistency cascade, we are in principle ready to tackle the problem of computing $\textrm{H}^0(s|\d)$. However, as mentioned at the beginning of this section, the BRST-BV formalism will allow us to tackle that problem \emph{backwards}. This means that, instead of classifying the $a_0$'s satisfying Equation \rf{cocycle3} and then working our way up the consistency cascade, we shall start with classifying the $a_2$ satisfying Equation \rf{cocycle1} and from there make progress all the way down to the corresponding, consistent $a_0$. Such is the power of the BRST-BV framework: we classify the consistent gauge-algebra deformations first and from there extract, by solving the consistency cascade (first for $a_1$ and then for $a_0$), the corresponding Lagrangian deformations. In this fashion the search for consistent deformations is rendered systematic and involves only the solving of precise cohomology equations. Also note that, by construction, a byproduct of this method is that the gauge-symmetry and gauge-algebra deformations corresponding to some found $a_0$ are readily available.\\

Let us discuss non-Abelian vertices first. The strategy is the following: classify all the $a_2$ satisfying Equation \rf{cocycle1}. Take a linear combination of all of them with arbitrary coefficients and plug it into Equation \rf{cocycle2}. Then solve Equation \rf{cocycle2} for $a_1$. Finally, plug the found $a_1$ into Equation \rf{cocycle3} and solve for $a_0$: it is the non-Abelian deformation of the Lagrangian. 

Actually, the classification of $a_2$ is further constrained by an equivalence relation. Indeed, two different $a_2$, both satisfying Equation \rf{cocycle1}, might yield the same $a_0$. This simply stems from the fact that, so far, in addressing the computation of $\textrm{H}^0(s|\d)$ we have only analyzed the condition that $a$ should be $s$-closed modulo $\d$, and in \rf{as} we have expanded it in antighost number. The condition of non-triviality in the cohomology should also be taken into account, and this means that the $a$ candidates are defined up to the equivalence relation given by the addition of $s$-exact terms modulo $\d$, that is, terms of the form $s m + \d n$. Now, as the reader shall easily verify upon recalling that $m$ and $n$ should also have total ghost number zero (and also stop at antighost number $2$), such an equivalence relation yields the following three equivalence relations for the different components of $a$:
\begin{subequations}
\vspace*{-10pt}
\label{equiv}
\begin{align}
a_2' &\sim a_2 + \Gamma b_2 \,+ \d c_2\,, \\
a_1' &\sim a_1 + \Delta b_1 + \d c_1\,, \\
a_0' &\sim a_0 + \Delta b_0 + \d c_0 \,.
\end{align}
\end{subequations}
This means that, when listing all the $a_2$ satisfying Equation \rf{cocycle1} we do so up to the above equivalence, and indeed one straightforwardly checks that two $a_2$'s differing by $\Gamma$-exact terms modulo $\d$ yield the same $a_0$, if any. Those would thus be two equivalent ways of writing down the gauge-algebra deformation induced by some given $a_0$. 

We now address Abelian vertices, that is, deformations $a$ for which the $a_2$ part is trivial, i.e. $\Gamma$-exact up to total derivatives. Now comes of use another theorem: when $a_2$ is trivial one can always choose it to be zero, and hence the consistency cascade starts one step lower, with Equation \rf{cocycle2} at $a_2 = 0$, that is, $\Gamma a_1 \doteq 0$ \cite{Barnich:1994cq}. Even better, the theorem further guarantees that one can chose $a_1$ to be exactly $\Gamma$-closed, and not only modulo $\d$. The Abelian vertices which nonetheless deform the gauge transformations are thus found by classifying all the inequivalent $a_1$'s which are $\Gamma$-cocycles. Again, two equivalent $a_1$'s, differing by $\Delta$-exact terms up to total derivatives, are seen to yield the same $a_0$. 

Last of all we address the `completely Abelian' vertices, namely those that not only preserve the gauge algebra but also leave the gauge transformations undeformed. This kind of deformations will have zero $a_2$ and trivial $a_1$, that is, the $a_1$ piece will be $\Delta$-exact modulo $\d$. In that case one can evidently remove $a_1$ so to be left with only Equation \rf{cocycle3} at $a_1 = 0$, to be solved for $a_0$, i.e. $\Gamma a_0 \doteq 0$, which is to be solved in light of the equivalence relation for $a_0$, simply given by field redefinitions and total derivatives. \\

One point worth highlighting is the crucial role played by the cohomology of $\Gamma$ at antighost number $2$, which is indeed the one computing the inequivalent $a_2$ candidates in the non-Abelian case.\footnote{Actually, as has been emphasized, one can choose $a_2$ to be strictly $\Gamma$-closed, and not only $\Gamma$-closed modulo $\d$.} ~\\

There is a subtlety in the non-Abelian case which we should comment on right away, and which we have overlooked so far in order not to crowd our first approach of the consistency cascade, but which is nonetheless an important point. Let us consider some $\Gamma$-closed $a_2$ and plug it into Equation \rf{cocycle2} in order to solve for $a_1$. The subtlety is the following: the solution for $a_1$, if it exists, is in fact defined up to $\Gamma$-closed terms only. Indeed, if two $a_1$'s differ by $\Gamma$-closed terms they will correspond to the same $a_2$. The solution $a_1$, if it exists, is then usually denoted as 
\be 
a_1 = \hat{a}_1 + \tilde{a}_1\,,
\ee
where the $\Gamma$-closed term $\tilde{a}_1$ is called the ambiguity and the non-ambiguous piece $\hat{a}_1$ is the solution found to solve Equation \rf{cocycle2} for our candidate $a_2$. Now comes the complication: when plugging the above $a_1$ into the last consistency equation, $\Delta a_1 + \Gamma a_0 \doteq 0$, it might be that a solution for $a_0$ only exists for some ambiguity $\tilde{a}_1$, and in general that is the case. In fact, this is the way the ambiguity is fixed. The computational intricacy is then that, in general, it might be difficult to either guess the correct ambiguity or express it into its most general form to then plug it into the last equation. For non-Abelian vertices the situation is thus usually the following: determining whether a candidate $a_2$ has a corresponding $a_1$ is not extremely difficult but, if there is such an $a_1$, finding out the correct ambiguity (or establishing that no $a_0$ solves the last equation, whatever ambiguity is used) can be tricky, and it is the most non-trivial part of the procedure. 

Finally, before addressing second-order consistency in the next step, let us comment on a procedural point, having to do with Abelian vertices. For the latter, the equation to solve is $\Gamma a_0 \doteq 0$, as we have just seen. However, it might not be the easiest one to solve and, as it usually turns out, it is easier to allow for $\Delta$-exact terms in $\Gamma a_0$. It is then easier to find the inequivalent $a_0$'s satisfying an equation of the form $\Gamma a_0 \doteq \Delta (\dots)$. The corresponding $a_1$ is then in general not equal to zero. However one can check that, because $a_1$ is trivial, it can always be canceled by the addition of $\Delta$-exact terms in $a_0$. By performing field redefinitions at the level of our vertex one can thus render its invariance manifest up to total derivatives only (or do the inverse thing). Differently put, depending on the chosen representation for our vertex, the absence of deformation of the gauge transformations (triviality of $a_1$) may appear explicitly ($a_1 = 0$) or not ($a_1 \neq 0$). 

\subsection*{Step 7: Second-Order Consistency and Antibracket}\addcontentsline{toc}{subsection}{Step 7: Second-Order Consistency and Antibracket}

Everything we have mentioned so far had to do with first-order consistency. In the completely Abelian case, when only the Lagrangian is deformed, no quartic or higher-order terms are needed and the consistency is automatic to all orders in perturbation theory. In the non-Abelian case where the gauge algebra is deformed (and in the `intermediate' case too) the situation is different; either the vertex is consistent to second order only up to the addition of a quartic term, $S_2$, or it is obstructed. In general, determining whether a non-Abelian vertex is obstructed or not and, in the latter case, determining the quartic term that needs to be added in order to render the theory fully consistent is rather tedious. It is however remarkable that the BRST-Antifield construction also provides one with just the right tool to deal with this issue, and that tool is called the \emph{antibracket}, which we now introduce. 

One defines the following odd, symplectic structure on the space of functionals of our fields and antifields:
\beq (X,Y)\equiv\frac{\d^RX}{\d\Phi^A}\frac{\d^LY}{\d\Phi^*_A}-\frac{\d^RX}{\d\Phi^*_A}\frac{\d^LY}{\d\Phi^A}\,.
\eeq{antibracket}
This definition gives $(\Phi^A,\Phi^*_B)=\delta^A_B$, which is real. Because a field and its antifield have opposite Grassmann parity, it follows that if $\Phi^A$ is real, $\Phi^*_B$ must be purely imaginary, and vice versa.\footnote{Recall that the complex conjugation acts as $(ab)^* = b^* a^*$, whose symbol `*' has nothing to do with the superscript distinguishing antifields.} Note that the antibracket satisfies the graded Jacobi identity. \\

To understand the usefulness of the antibracket, we first note the following peculiar and a priori anodyne fact: the action of the free BRST differential $s$ can be rephrased as taking the antibracket with the free master action $S_0$, that is,
\be 
s F = (S_0,F) \,,
\ee
for any phase-space functional $F$. Better yet, one can actually prove the equivalent of the above statement for the fully deformed theory too\,! Let us denote the completely deformed BRST differential by $\mathfrak{s}$, so that\footnote{Note that, to be homogeneous in our use of notation we should have called the full BRST operator $s = s_0 + s_1 + \cdots$, but as the zeroth-order part is the most often used piece we have chosen to be more economical.}
\be 
\mathfrak{s} \equiv s + s_1 + s_2 + \cdots \,,
\ee
where for example $s_1$ is the sum of some $\Gamma_1$, implementing the deformed piece brought in by $a_1$, and some $\Delta_1$ which implements the contribution to the free equations of motion induced by $a_0$. The full statement is then 
\be 
\mathfrak{s} F = (S,F) \,,
\ee
which can be seen to hold by virtue of the Noether identities and the higher-order gauge-structure equations \cite{Henneaux:1989jq}. 

We are now ready to formulate an equation which is the cornerstone of the BRST-BV approach to Gauge Theory. Indeed, the full master action $S$ is invariant under the full BRST differential $\mathfrak{s}$, so that by virtue of the above relation we have
\be 
\label{masterequation}
(S,S) = 0 \,.
\ee
This is the so-called (classical) \emph{master equation}, which contains all the information about Noether identities and higher-order gauge structure equations. It remarkably repackages all the conditions defining a fully consistent deformation into a single `geometrical' equation. This structure allows for a rephrasal of many a property. For example, in this way one can see the nilpotency of $\mathfrak{s}$ as a mere consequence of the graded Jacobi identity for the antibracket. Furthermore, let us also point out that the above odd structure is related to the more familiar Poisson bracket, and to other structures as well \cite{Barnich:1996mr}, but in the present guide we shall not dwell on these interesting questions. 

To see how this new structure helps addressing the problem of second-order consistency let us split the master equation above in terms of the coupling constant $g$ by inserting in it the perturbative expression \rf{masteractionpert}. The first orders give us
\begin{subequations}
\begin{align}
(S_0,S_0)&= 0\,,\label{brst6.1}\\
(S_0,S_1)&=0\,,\label{brst6.2}\\
(S_1,S_1)&=-2(S_0,S_2)\,.\label{brst6.3}
\end{align}
\end{subequations}
The first equation here above is satisfied by assumption: it can be rewritten as $s S_0 = 0$, which is simply the statement of invariance of the free master action under the (free) BRST differential. The second equation translates to $s S_1 = 0$, which is the integrated version of the cohomological condition written down in \rf{acohom}. As for the third one, it expresses in a compact way the condition that $S_1$ must satisfy so as to be consistent at second order, where it is completed by a quartic term $S_2$. It determines whether or not, in a local theory, a consistent first-order deformation gets obstructed
at the second order. One thus sees that second-order consistency is controlled by the local cohomology group $\textrm{H}^1 (s)$, for $(S_1,S_1)$ has ghost number $1$. More precisely, one easily checks that $(S_1,S_1)$	is BRST-closed (by the graded Jacobi identity), and what the third equation here above does is to further require it to be trivial in $\textrm{H}^1 (s)$. Keeping in mind that $s$ annihilates $(S_1,S_1)$ one may thus reexpress the requirement of second-order consistency as
\be 
(S_1,S_1) \notin \textrm{H}^1 (s)\, .
\ee
Moreover it can be shown that $\textrm{H}^1 (s)$ is also the cohomology group controlling higher-order deformations. However, more often than not in Gauge Theory a deformation is either fully consistent, that is, to all orders, or consistent only at the cubic level, and hence fails to satisfy the above requirement. Note that in the above condition it is truly the cohomology of $s$ which is used, not the cohomology modulo $\d$, so that one should expect strong conditions to arise from it. \\

There is (even) more to the antibracket. Indeed, with such a symplectic structure one can actually reformulate the whole problem of deformation of the free master action. As we have seen the free master action does satisfy the master equation \rf{masterequation}, and the full deformation should also fulfill it. The problem of consistently deforming a free theory can thus be reformulated as the problem of deforming the solution $S_0$ to the master equation, and this allows for a different mathematical approach to the problem, which has proved very useful \cite{Gerstenhaber:1964zz}. More generically speaking, the BRST-BV reformulation presented here has allowed for a systematic study of many aspects of Gauge Theory, as for example that of \cite{Boulanger:2000rq}, where the above techniques are used to discard as inconsistent theories involving more than one graviton in interaction and the unicity of the Einstein--Hilbert action is proved under very generic assumptions. 

We end this section with a reminder of the quantum numbers for our fields and antifields as well as the action of the different operators on them. In the free, irreducible case of interest to us they read as follows.  
\begin{table}[ht]
\label{table:summaryrecap}
\caption{Properties of the Various Fields, Antifields and Operators}
\vspace{5pt}
\centering
\begin{tabular}{c c c c c c c}
\hline\hline\vspace*{-10.5pt}\\
$Z$ &$\Gamma(Z)$&$\Delta(Z)$&$\pgh(Z)$ &$\agh(Z)$ &$\gh(Z)$ &$\epsilon(Z)$\\ [0.2ex]
\hline\vspace*{-10.5pt}\\
$\phi^i$ & $R^i_\a \mathcal{C}^\a$ & 0 & 0 & 0 & 0 & 0\\
$\mathcal{C}^\a$ & 0 & 0 & 1 & 0 & 1 & 1\\
$\phi^*_i$ & 0 & $\textrm{E}_i[\phi^j]$ & 0 & 1 & $-1$ & 1\\
$\mathcal{C}^*_\a$ & 0 & $R^i_\a \phi^*_i$ & 0 & 2 & $-2$ & 0 \\\vspace*{-10.5pt}\\
\hline\hline
\end{tabular}
\end{table}

\section{Quantum Electrodynamics: a Simple Example}
\label{sec:qed}

The BRST-Antifield reformulation of the interaction problem has been introduced in the previous section for free, irreducible theories. Let us now illustrate these techniques on the simple example of quantum electrodynamics. To be fair we should point out that the this example is perhaps a little treacherous, in the sense that there is no gauge invariance for the fermion and hence the problem of building consistent interactions becomes much, much simpler as we shall see. We believe it is nevertheless a good place to start applying the formalism. 

Let us construct all the off-shell cubic vertices involving a particle of spin~$1$ and a particle of spin~$\tspin$. Our assumptions are Lorentz invariance, Parity invariance and locality. Lorentz invariance means that spacetime indices must be contracted with one another using the Minkowski metric $\eta_{\mu\nu}$ (see Section~\ref{context}), and Parity invariance boils down to foregoing the use of an odd number of epsilon symbols. The starting point is the free theory, which thus contains a photon $A_\mu$ and a massless electron field $\psi$,
described by the action
\beq  S^{(0)}[A_\mu,\psi]=\int\tintspace \d^\tdim x\left(-\tfrac{1}{4}F_{\mu\nu}^2-i\bar{\psi} \ds\psi\right)\,,
\eeq{qed1} 
which enjoys the Abelian gauge invariance 
\beq \delta_\lambda A_\mu=\partial_\mu
\lambda\, ,\eeq{qed2} 
and no gauge invariance for $\psi$. In the above free action $\bar{\psi}$ is the Dirac adjoint of $\psi$, $\bar{\psi}  \equiv \psi^\dagger \gamma^0$, where $\gamma^\mu$ are the Dirac gamma matrices with $\mu = 0,\dots,D-1$ and our notation goes $\not{\!Q} \equiv  \gamma^\mu Q_\mu$ for any vector $Q_\mu$. Our convention for the gamma matrices is $\{\gamma^\mu,\gamma^\nu\} = 2\eta^{\mu\nu}$ and $\gamma_\mu^\dagger = \eta_{\mu\mu}\gamma_\mu = \gamma^0 \gamma_\mu \gamma^0$. The conjugation operator $\bullet^\dagger$ is an involution which squares to the identity and determines the reality properties of the various objects under consideration. Our convention regarding reality conditions is the following: $A_\mu^\dagger = A_\mu$, while $\psi$ is complex and conjugated to $\psi^\dagger$. As is easily checked, the above master action is hermitian. 

For the Grassmann-even bosonic gauge parameter $\lambda$ we introduce the Grassmann-odd bosonic ghost $C$, and no ghost corresponding to $\psi$ is introduced for the latter enjoys no gauge invariance.\footnote{Here is where our setup is a little misleading in illustrating the BRST-BV methods, for only one ghost needs to be introduced which will drastically simplify certain aspects.} Therefore the set of fields becomes
\beq \Phi^A=\{A_\mu, C, \psi \}\,.\eeq{qed3} For each of these fields, we introduce an antifield with the same algebraic
symmetries in its indices but opposite Grassmann parity. The set of antifields thus is \beq \Phi^*_{A}=\{A^{*\mu}, C^*, \bar{\psi}^{*}
\}\,.\eeq{qed4} Now we construct the free master action $S_0$, which is an extension of the original
gauge-invariant action~(\ref{qed1}) by terms involving ghosts and antifields. Explicitly, \beq S_0=\int\tintspace \d ^\tdim x\left(-\tfrac{1}{4}
F_{\mu\nu}^2-i\bar{\psi} \ds\psi + A^{*\mu}\partial_\mu C\right)\,.\eeq{qed5} Notice how the antifields appear as sources
for the gauge variations, with gauge parameters replaced by corresponding ghosts. It is easy to verify
that the above free action $S_0$ indeed solves the master equation $(S_0,S_0)=0$. The different gradings and Grassmann parity
of the various fields and antifields, along with the action of $\Gamma$ and $\Delta$ on them, are given in Table~$2$ below. 

\begin{table}[ht]
\label{tablenXX}
\caption{Properties of the Various Fields \& Antifields}
\vspace{5pt}
\centering
\begin{tabular}{c c c c c c c}
\hline\hline\vspace*{-11pt}\\
$Z$ &$\Gamma(Z)$&$\Delta(Z)$&$\pgh(Z)$ &$\agh(Z)$ &$\gh(Z)$ &$\epsilon(Z)$\\ [0.2ex]
\hline\vspace*{-10.5pt}\\
$A_\mu$ & $\partial_\mu C$ & 0 & 0 & 0 & 0 & 0\\
$C$ & 0 & 0 & 1 & 0 & 1 & 1\\
$A^{*\mu}$ & 0 & $-\partial_\nu F^{\mu\nu}$ & 0 & 1 & $-1$ & 1\\
$C^*$ & 0 & $-\partial_\mu A^{*\mu}$ & 0 & 2 & $-2$ & 0 \\\vspace*{-11pt}\\ \hline\vspace*{-11pt} \\
$\psi$ & 0 & 0 & 0 & 0 & 0 & 1\\
$\bar{\psi}^{*}$ & 0 & $-i\bar{\psi}\cev{\ds}$ & 0 & 1 & $-1$ & \;0 \vspace*{0.5pt}\\
\hline\hline
\end{tabular}
\end{table}

The cohomology of $\Gamma$ is isomorphic to the space of functions of
\begin{itemize}
 \item The undifferentiated ghost $C$,
 \item The antifields $\{A^{*\mu}, C^*, \bar{\psi}^{*}\}$ and their derivatives,
 \item The curvature $F_{\mu\nu}$ and its derivatives,
 \item The field $\psi$. 
\end{itemize}

Let us now classify the consistent cubic couplings. We start with the non-Abelian ones. In fact, it is easily seen that there can be no consistent non-Abelian couplings involving both types of fields in the present setup. Indeed, the construction of a candidate $a_2$ involving either $\psi$ or $\bar{\psi}^{*}$ fails at the level of the quantum numbers already, as can be derived by looking at the above table and further recalling that $a_2$ must be a cubic combination of total ghost number zero and antighost number two. From those considerations only, one sees that any $a_2$ should be of the schematic form $\textrm{ghost} \times \textrm{ghost} \times \textrm{antighost}$, and as we have no ghost corresponding to $\psi$ (because it enjoys no gauge invariance) one can only construct self-coupling $a_2$ candidates for $A_\mu$. We are not interested in those (which lead to the familiar Yang--Mills cubic term \cite{Barnich:1993pa} when the photon is colored), and wish to look at cross couplings only. The non-Abelian case is thus covered, and there are no non-Abelian consistent couplings. \\

Let us now address the Abelian couplings. We first investigate vertices which do not deform the gauge algebra but nevertheless deform the gauge transformations, and then move on to `completely Abelian' ones, namely those vertices which deform only the Lagrangian. In order to search for couplings with trivial $a_2$ but non-trivial $a_1$, let us classify the possible deformations of the gauge transformations. They should have antighost number $1$ and total ghost number zero, and further be cubic in our fields and antifields. Evidently, it should also be Lorentz invariant and have all spinor indices contracted. The only such combination with no derivatives is easily concluded to be
\be 
a_1 = g\bar{\psi}^* \psi C \,.
\ee
If such a gauge-symmetry deformation indeed does not deform the gauge algebra, it should satisfy $\Gamma a_1 \doteq 0$ (see previous section), which indeed it does. If it corresponds to a vertex, it must also be such that $\Delta a_1 + \Gamma a_0 \doteq 0$. One easily realizes that, only when the coupling constant $g$ is imaginary does the above deformation get lifted to a consistent Lagrangian vertex. Indeed, making use of partial integration the cohomology equation is easily solved for $a_0$, which is found to be
\be 
\label{vertexonly}
a_0 = ig\bar{\psi}\As\psi\, .
\ee 

One should now investigate the fate of $a_1$ candidates containing derivatives. However, those are immediately ruled out as trivial. To see it, let us first make clear that the antifield in $a_1$ can always be assumed to be undifferentiated, as $a_1$ is defined up to total derivatives only. Now, if a derivative acts on the ghost, it would produce the gauge variation of $A_\mu$, which is by definition a $\Gamma$-exact object, and because $\Gamma$ does not act on $\psi$ nor $\psi^*$, this situation would correspond to a $\Gamma$-exact $a_1$ (one can pull out $\Gamma$ to make it act on the whole $a_1$ above), and this would correspond to a $\Delta$-exact $a_0$, which is trivial (see Step~$6$ of the previous section). If, on the other hand, a derivative acts on $\psi$, the following argument can be used: this derivative cannot come alone (because of Lorentz invariance), and there are no indices on the involved fields, so that it must be contracted either with another derivative or with a Dirac $\gamma$-matrix. Because no derivatives can act on the ghost (see above) and because $\square = \ds \ds$, both these situations give rise to the equations of motion in $a_1$, that is, to $\Delta \psi^* = -i \psi^*\cev\ds$ or combinations thereof. This does not generically make the whole deformation $\Delta$-exact, because the $\Delta$-operator acts on $\bar{\psi}^*$, but one can check that the cohomology equation $\Delta a_1 + \Gamma a_0 \doteq 0$ is then either not satisfied ($g\in \mathbb{C}\tintspace\setminus\tintspace\mathbb{R}$), or it is satisfied ($g\in \mathbb{R}$) but the resulting $a_0$ is $\Delta$-exact. Considering even more derivatives only makes the situation worse, and we thus conclude that there is only one vertex which deforms the gauge transformations and it has zero derivatives. 

Our search is now narrowed down to the couplings which preserve the gauge transformations. The only part of the deformation that we need care about is thus $a_0$, and the $a_1$ piece can always be chosen to be zero (see previous section). We are left with the equation $\Gamma a_0 \doteq 0$ to be solved, and we shall alternatively use the weaker equation $\Gamma a_0 \doteq \Delta (\dots)$. We start with vertices containing no spacetime derivatives. We directly see that the only Lorentz-invariant possibility is
\be 
a_0 = \bar{\psi} \As \psi \,,
\ee
which obeys $\Gamma a_0 \doteq \Delta (\dots)$. This is the vertex we have already found above. Note that, if we had not found this vertex previously, we could be tempted to conclude that the latter is completely Abelian. However, because we have used the weaker equation $\Gamma a_0 \doteq \Delta (\dots)$ here, this is not guaranteed, and in this instance it is of course not true. We take this opportunity to recall that, when using the weaker equation one should check whether the corresponding $a_1$ is trivial, that is, whether it can be canceled by field redefinitions at the level of the Lagrangian. 

We then address the vertices containing one derivative. An obvious possibility is the term built in terms of the curvatures (in this case there is only one curvature, namely that of $A_\mu$, for the fermion has no spacetime indices):
\be 
\label{prodcurv}
a_0 = \bar{\psi} \gamma^\mu \gamma^\nu F_{\mu\nu}\psi \,,
\ee
which is strictly gauge invariant (not even modulo $\d$). The only other Lorentz-invariant combination with one derivative is $\partial^\mu A_\mu \,\bar{\psi} \psi$, but it is easily seen to violate the consistency equation $\Gamma a_0 \doteq 0$. Furthermore, it is easily proved that there are no higher-derivative candidates, for all such Lorentz-invariant combinations would be on-shell trivial up to partial integration, as the reader shall easily convince himself of. \\

Let us comment on the nature of the vertices. The last one \eqref{prodcurv} is a product of curvatures and is strictly gauge invariant. The other one is seen to be different: whatever field redefinition we perform on it the best we can do is bring it to a form in which it is on-shell gauge invariant modulo $\d$. That vertex is also the one which completes the free kinetic term for the fermion, turning it into the familiar expression involving the covariant derivative $\bar{\psi} \Ds \psi$, with $D_\mu \equiv \partial_\mu - igA_\mu$. The said cubic coupling is thus the one resulting from covariantizing the derivatives in the fermion kinetic term, namely, the so-called minimal coupling, which in the present setup has zero derivatives. Although simple, for future comparison it is useful to summarize the results of the present section, and we give them in the table hereafter, where $\tspin$-Abelian means that the gauge transformations are deformed but in a way which preserves the original (Abelian) gauge algebra of the free theory. 
\vspace{-5pt}
\begin{table}[ht]
\label{electronem}
\caption{Summary of $\vertex{1}{\tspin}$ Vertices}
\vspace{5pt}
\centering
\begin{tabular}{c c c c}
\hline\hline\vspace*{-10.5pt} \\
$\#$ of derivatives~~~&~~~Vertex~~~~~&~~~~~Nature~~~~~\\ [0.2ex]
\hline\vspace*{-10.5pt}~\\
0 & $\bar{\psi} \As \psi $ & $\tspin$-Abelian \\
1 & $\bar{\psi} \gamma^\mu \gamma^\nu F_{\mu\nu}\psi$ & Abelian \\\vspace*{-10.5pt}~\\
\hline\hline\vspace{0pt}
\end{tabular}
\end{table} 

As mentioned above, the example treated here fails to capture some of the complexity of the deformation problem, for there is not deformation of the gauge algebra, which drastically simplifies the classification. A less trivial exercise which remains easy to go through is that of Yang--Mills Theory, which is analyzed in \cite{Barnich:1993pa} by means of the BRST--Antifield technique. 

\section{Final Remarks}

In this short and direct review we have attempted at introducing the BRST-Antifield reformulation of the interaction problem in a pedagogical fashion, emphasizing its physical meaning and focusing on the simplest case where the starting theory is free in a Minkowski spacetime of arbitrary dimension $D\geq 4$. As we have tried to convince the reader, the reformulation of the problem of finding consistent (gauge-invariant) and non-trivial vertices starting from some free Lagrangian using the BRST--BV formalism is most helpful. Despite its apparent complexity, once the framework is constructed one gains many advantages: off-shell and invariant cubic vertices can be obtained and classified in a systematic and exhaustive way, the induced deformations of the gauge transformations and gauge algebras are automatically obtained, field redefinitions are well under control and the question of second-order consistency can be most elegantly addressed by making use of the antibracket. In brief, we argue that the reformulation of a physics problem into such a sharp, cohomological question grants one a clearer understanding of the precise way the physical theory is being deformed by the vertices, and allows one to classify the latter in a clever fashion: starting from inequivalent, putative gauge-algebra deformations and arriving at the Lagrangian ones, not the other way around. The literature on the BRST-Antifield reformulation of the deformation problem includes the very good review \cite{Henneaux:1997bm} and we also point out the algebraic and geometrically oriented lectures \cite{Henneaux:1989jq}, the report \cite{Barnich:2000zw} as well as the comprehensive book \cite{henneauxbook}, all of which go beyond the scope of the present review. 

Before putting an end to our straight story, let us make some comments on settings we have not discussed and hypothesis we have made. First let us make clear once again that, although we have addressed the case where the starting point is a free theory, the formalism we have presented can be extended to reformulate the problem of deforming any theory\;---\;with some modification.\footnote{More precisely, the BRST differential will contain additional pieces when at least one of the following properties characterizes the starting theory: the gauge symmetries are not Abelian or they form an \emph{open} algebra \cite{Henneaux:1989jq}.} In the case at hand we have taken advantage of the fact that the gauge transformations we start with are irreducible, which is not necessarily the case for an interacting theory. Secondly we point out that, in principle, the BRST-BV framework does not rely on the spacetime manifold being Minkowskian and can be applied to a theory formulated on any background. However this could suppose practical difficulties, as for example in anti-de Sitter where the explicit dependence of the free Lagrangian on the spacetime coordinates (i.e. not only via the fields) makes things less obvious.
Also we note that, even though the case we have explicitly worked out in the previous section deals with vector bosons and spin-$1/2$ fermions, the formalism is not restrictive in this sense and can be used to study systems describing any kinds of field tensors. Works making use of the Antifield formalism include \cite{Boulanger:2001wq,Boulanger:2006gr,Boulanger:2008tg,Bekaert:2010hp,Fotopoulos:2007yq,Henneaux:2012wg,Henneaux:2013gba}.\footnote{The present introduction is heavily based on Chapter~$4$ of the author's PhD thesis \cite{Gomez:2014dwa}, where the formalism is then employed in order to classify higher-spin couplings to Electromagnetism and to Gravity.} Finally, let us stress once again that locality of the would-be consistent couplings is assumed through the text. In fact, one can show that if locality is not insisted on then any consistent cubic coupling can be completed to the quartic order \cite{Barnich:1993vg}. \\

We hope the present guide has somewhat helped the reader grasping the meaning of the BRST-Antifield reformulation of the deformation problem as well as its elegance. 

\section*{Acknowledgments}

It is a pleasure to warmly thank Jim~\textsc{Stasheff}, not only for his many valuable comments during the preparation of the draft but also for the incentive for publishing the present review. During the final stages of the writing process the author has benefited from insightful and most helpful remarks by Nicolas~\textsc{Boulanger}, Micha~\textsc{Moskovic}, Rakibur~\textsc{Rahman} and Florian~\textsc{Spinnler}, which we are most thankful for. Finally, the author's understanding of the material which is presented here would have been impossible without the guiding and patience of Marc~\textsc{Henneaux} and Rakibur~\textsc{Rahman}. The author is a Fellow of the Alexander von Humboldt Foundation and his work was also supported by the Grant Agency of the Czech Republic under the grant 14-31689S.

		\sloppy
	\printbibliography[resetnumbers=true] \addcontentsline{toc}{section}{References}%
\end{document}